\documentclass{PoS}

\newif\ifgeneratebbl
\generatebbltrue

\usepackage{epsfig}
\usepackage{amsmath,amssymb}
\usepackage[squaren]{SIunits}

\renewcommand{\thesection}{\arabic{section}}

\newcommand{\eq}[1]{Eq.~(\ref{#1})}
\newcommand{\fig}[1]{Fig.~{\ref{#1}}}
\newcommand{\be}{\begin{equation}}
\newcommand{\ee}{\end{equation}}
\newcommand{\bea}{\begin{eqnarray}}
\newcommand{\eea}{\end{eqnarray}}
\newcommand{\DS}{Dyson--Schwinger~}
\newcommand{\RG}{renormalization~group~}
\newcommand{\w}{\omega}

\newcommand{\al}{\alpha}
\newcommand{\ba}{\beta}
\newcommand{\ga}{\gamma}

\newcommand{\de}{\delta}

\newcommand{\La}{\Lambda}

\newcommand{\pd}{\partial}

\newcommand{\cs}{{\cal S}}

\newcommand{\im}{\mathrm{i}}

\newcommand{\prb}{Phys.\ Rev.\ B}
\newcommand{\prd}{Phys.\ Rev.\ D}

\title{Effects of electron-electron interactions in suspended graphene}

\ShortTitle{Effects of electron-electron interactions in suspended graphene}

\author{\speaker{C.~Popovici}\\
 Institut f\"ur Theoretische Physik, Justus-Liebig-Universit\"at Giessen
 Heinrich-Buff-Ring 16, 35392 Giessen, Germany  \\
        E-mail: \email{carina.popovici@theo.physik.uni-giessen.de}}

\author{{C.~S.~Fischer}\\
Institut f\"ur Theoretische Physik, Justus-Liebig-Universit\"at Giessen
 Heinrich-Buff-Ring 16, 35392 Giessen, Germany }

\author{{L.~von~Smekal}\\
Theoriezentrum, Institut f\"ur Kernphysik, Technische Universit\"at
Darmstadt,\\  
Schlossgartenstra{\ss}e 2, 64289 Darmstadt, Germany}

\abstract{
We investigate the problem of dynamical gap generation in suspended
graphene by long-range Coulomb interactions at strong coupling with
Dyson-Schwinger equations. Including renormalization effects on the
Fermi velocity we obtain a critical coupling constant $\alpha_c=2.85$
which is larger than the bare coupling $\alpha_0=2.19$ of suspended
graphene. This suggests that at low energies the running of the Fermi
velocity prevents the emergence of a gapped phase. Our calculation is
thus in agreement with the experimental observation that suspended
graphene remains in the semimetal phase.}

\FullConference{Xth Quark Confinement and the Hadron Spectrum\\
                 8-12 October 2012\\
                 TUM Campus Garching, Munich, Germany}

\begin{document}

%--------------------------------------------------------------
\section{Introduction}

Due to its unusual electronic properties, graphene, a monolayer of
carbon atoms on a two-dimensional honeycomb lattice
\cite{Novoselov2004}, has been intensely studied both theoretically
and experimentally over the last years (see, for example,
Refs.~\cite{kouc2008,CastroNeto:2009zz,Kotov:2010yh} for overviews).  In
particular, much work has been devoted to understanding the effects of
electron-electron interactions, both in the continuum, mainly via \DS
and \RG equations
\cite{Gonzalez94,Gonzalez2001,jugr2010,gogu2001,gogu2002,gago2010,
khve2001,khve2009, Mesterhazy:2012ei,druso2008,herbut2006}, and on the
lattice \cite{drla12009,drula2009,Drut:2009zi,
Buividovich:2012nx,Buividovich:2012uk}.  The majority of the
theoretical efforts were thereby focussed on the question whether the
long-range Coulomb interactions might trigger the dynamical generation
of a gap in graphene at strong coupling similar to chiral symmetry
breaking in QCD. This possibility is motivated by the potentially
large value of the effective coupling $\al=e^2/\varepsilon\hbar v_F$,
where $v_F\sim 10^6\, \mathrm{m/s}\approx c/300 $ is the Fermi
velocity of graphene on a substrate with dielectric constant
$\varepsilon $.  If the Fermi velocity of suspended graphene with
$\varepsilon =1$ is of the same order, this could lead to the
formation of a condensate corresponding to a transition into an
insulating phase above some critical coupling $\al_c$.

In a series of theoretical works, various authors have calculated the
value of the critical coupling that would correspond to the
semimetal-insulator transition: Gamayun \emph{et al.}~found
$\alpha_c=0.92$ by solving a simplified gap equation in which
radiative corrections to the Fermi velocity are neglected
\cite{gago2010}; in \cite{khve2009}, an approximation for the photon
polarization has been used, leading to the critical value
$\alpha_c=1.13$; renormalization group calculations at two loops yield
$\alpha_c=0.833$ \cite{vafek2008}; on the lattice, calculations
performed using a 'standard' square lattice found the value
$\al_c=1.08\pm 0.05$ \cite{drula2009}, whereas simulations carried out
directly on a hexagonal (physical) lattice yield $\al_c=0.9\pm 0.2$
\cite{Buividovich:2012nx}. All these values lie below the bare
coupling constant of suspended graphene, $\al_0=2.19$.   

On the other hand, the experimentally measured conductivity shows that
suspended graphene is not in the insulating phase
\cite{Kotov:2010yh,elgo2012}.  It is observed, however, that the strong
Coulomb interactions induce a charge-carrier density dependent
renormalization of the Fermi velocity corresponding to a reshaping of
the Dirac cones.  This renormalization peaks at half-filling where it
leads to an increase by about a factor of three in the Fermi velocity
and thus a corresponding decrease in the renormalized effective
coupling. A strong renormalization of $v_f$ was predicted already in
an early perturbative RG study \cite{Gonzalez94} where it was
concluded that the Coulomb interactions would in fact induce a
logarithmic increase in the Fermi velocity without bound until
retardation effects eventually become important and the instantaneous
Coulomb approximation breaks down.

While lattice simulations can yield reliable results in strongly
coupled theories, extrapolations to the infinite volume and chiral
limits are often a numerical challenge. This is also very much true
for the simulations of the electronic properties of graphene.
Therefore, functional methods in which these limits are accessed more
easily can provide important alternative nonperturbative formulations.
Moreover, the inevitable approximations in these formulations can be
validated from simulations with sufficiently large masses in finite
volumes. Here we employ the \DS equations. Within this approach,
substantial progress has been made over the past decades, both in in
QCD and QED$_3$ (see, for example,
Refs.~\cite{Alkofer:2000wg,Maris:2003vk,Fischer:2006ub,Maris:1995ns,fial2004}
and references therein). Extending the study of Ref.~\cite{gago2010},
we first calculate the running Fermi velocity in the bare vertex
approximation and with a one-loop (frequency dependent) photon
polarization. In a second step, we solve the gap equation at the
critical point by explicitly including the dressing function
corresponding to this running Fermi velocity. For the critical
coupling of the semimetal-insulator transition we then obtain
$\al_c=2.85$ \cite{inprep} which is larger than the bare coupling of
suspended graphene $\al_0=2.19$, indicating the persistence of the
semimetal phase. Our results thus agree with the experimental findings
\cite{elgo2012}. Finally, in order to compare our results with the
lattice simulations which must include a sufficiently large mass term
in a finite volume, we have also investigated the dependence of the
corresponding pseudocritical coupling on such a fermion mass term
which acts as an explicitly symmetry breaking external field as the
current quark masses do in QCD.

%-----------------------------------------------------------------------------

\section{Running Fermi velocity and critical point analysis}
\setcounter{equation}{0}

We start by briefly sketching the details of the continuum model of
graphene employed in this work.  At the level of the Lagrangian,
including the Coulomb interaction corresponds to accommodating the
boson field that propagates in (3+1) dimensions into the (2+1)
dimensional Lagrangian of the massless Dirac fermions. This is done
via dimensional reduction, which gives rise to a so-called reduced
gauge theory (we refer to Refs.~\cite{gogu2002,gago2010,inprep} for a
complete description and derivation). The dimensionally reduced action
of 2D Dirac fermions interacting through a three-dimensional gauge
field is given by
\bea
\cs&=&\int dt d^2\vec r\,\, \bar\psi(t,\vec r)\left[\im\ga^0\pd_t-\im
  v_F\ga^i\pd_i\right]\psi(t,\vec r)
\nonumber\\
&&-\frac{e^2}{2\varepsilon}\int dt dt^\prime d^2\vec r  d^2\vec
r^{\,\prime}\, \delta(t-t^\prime) \bar\psi(t,\vec r)\ga^0\psi(t,\vec r) 
\frac{1}{|\vec r-\vec  r^{\,\prime}|}
\bar\psi(t^\prime,\vec r^{\,\prime})\ga^0\psi(t^\prime,\vec r^{\,\prime}).
\label{eq:braneaction}
\eea
Here, the spatial index is $i= 1,2$, and $s=1,...,N_f$ is a flavor
quantum number with $N_f=2$ for monolayer graphene referring to the
physical spin degree of freedom.  $\psi (\bar\psi)$ denotes the
fermion (conjugate) field, $v_F$ is the Fermi velocity, and
$\varepsilon$ is the dielectric constant of the medium (for suspended
graphene $\varepsilon =1$, whereas on top of a substrate one uses
$\epsilon = (1+\kappa)/2$ typically with $\kappa \approx 4$ for
SiO$_2$ or $\kappa \approx 10$ for SiC).

From the action in \eq{eq:braneaction} one obtains the fermion \DS
(gap) equation via standard functional derivation. In Minkowski space
we have
\be
S^{-1}(p_0,\vec p)=S_0^{-1}(p_0,\vec p)-\im e\ga^0\int
\frac{d^3k}{(2\pi)^3} S(k_0,\vec k)\Gamma(k,-p)D(p_0-k_0,\vec p-\vec k)\,,
\label{eq:DS}
\ee
where $\Gamma(k,-p)$ is the fully dressed fermion-photon vertex and
the fermion propagator is given by (without wavefunction
renormalization, details will be provided elsewhere \cite{inprep}):
\be
S(p)=\frac{\ga^0p_0-v_F A(p)\,\ga^ip_i-\Delta(p)}
{p_0^2-v_F^2A ^2(p)\,\vec p^{\,2}-\Delta^2(p)},
\ee
where $A$ is the Fermi velocity dressing function, and $\Delta$ is the
fermion mass function. The Coulomb propagator in the random-phase
approximation (RPA),
\be\label{one-loop}
D(q_0,\vec q)=\frac{2\pi}{|\vec q|+\Pi(q_0,\vec q)},
\ee
includes polarization effects with the one-loop polarization function
\cite{Gonzalez94},
\be 
\Pi(q_0,\vec
q)=\frac{\pi e^2 N_f}{4\varepsilon}\frac{\vec q^{\,2}}{\sqrt{\hbar^2
    v_F^2\vec q^2-q_0^2}}.
\ee
Following the usual procedure, we recast the \DS equation \eq{eq:DS}
into a set of coupled integral equations for the dressing functions
$A(p)$ and $\Delta(p)$. Neglecting a possible weak energy dependence
in $A$ and $\Delta$, and in a bare vertex truncation $\Gamma
\rightarrow \Gamma^{(0)}=e\ga^0$, one obtains \cite{gago2010}:
\bea
A(\vec p)
&=&1+\frac{e^2}{\varepsilon}\int_{-\infty}^{\infty}\frac{d\w}{2\pi}\int\frac{d^2\vec
  k}{(2\pi)^2}
\frac{\vec p\cdot \vec k}{\vec p^{\,2}}\frac{A(\vec k)}{\w^2+v_F^2
  A^2(\vec k)\vec
  k^2+\Delta^2(\vec k)}D(\w, \vec p-\vec k),\label{eq:gapA}\\
\Delta(\vec p)
&=&\frac{e^2}{\varepsilon}\int_{-\infty}^{\infty}\frac{d\w}{2\pi}\int\frac{d^2\vec
  k}{(2\pi)^2}
\frac{\Delta(\vec k)}{\w^2+v_F^2 A^2(\vec k)\vec
  k^2+\Delta^2(\vec k)}D(\w, \vec p-\vec k).\label{eq:gapD}
\eea 
In order to evaluate these equations at the critical coupling we apply
bifurcation theory \cite{atbl1994}, which amounts to linearizing
Eqs.~(\ref{eq:gapA}, \ref{eq:gapD}) around the point at which the
nontrivial solution for the mass function bifurcates away from the
trivial one.  In this case the integral equation for $A$ decouples.
Furthermore employing a GW approximation as in
Ref.~\cite{gonzalez1999} amounts to setting $A = 1$ on the r.h.s.~of
\eq{eq:gapA} such that the corresponding equation is reduced to
solving a single integral. The integration over the (Euclidean)
frequency $\omega$ can then be performed analytically \cite{gago2010},
yielding (now with $p\equiv |\vec p|$):
\begin{figure}[t]
%\vspace{0.5cm}
\centering
\includegraphics[width=0.6\linewidth]{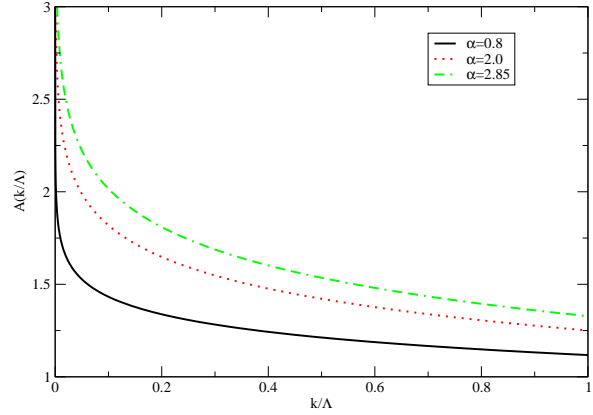}
\caption{\label{fig:A-crit-pct}
The Fermi velocity dressing function $A$ in GWA with RPA Coulomb
interactions, $\Lambda \sim \hbar/a $, $N_f=2$ for monolayer graphene,
and $\alpha = 0.8$, $2.0$ and $2.85$ to represent the presence of a
substrate, suspended graphene, and criticality for the
semimetal-insulator transition, respectively, from bottom to top.}
\end{figure}
\be
A(p)=1+\frac{e^2}{v_F\varepsilon} \frac{1}{\vec p^{\,2}}\int\frac{d^2\vec
  k}{(2\pi)^2}\frac{\vec p\cdot \vec k}{ k\,|\vec p-\vec
  k|}J\left(z,g\right),\label{eq:gapA2}
\ee
where $g= e^2 N_f \pi/(4\varepsilon v_f)$ and the piecewise function
$J(z,g)$ with $z\equiv |\vec k|/|\vec p-\vec k|$ is given by 
\bea
J(z,g)=\frac{(z^2-1)[\pi-g c(z)]+zg^2 c(g)}{z^2+g^2-1},
\mathrm{~~with~~}
c(x)=\left\{ 
  \begin{array}{l l l}
2\,\mathrm{arccosh~}(x)/\sqrt{x^2-1} &, & \; x>1\vspace{1mm}\\
2\arccos (x)/\sqrt{1-x^2}&, & \; x<1\vspace{1mm}\\
2&, & \; x=1
  \end{array} \right. .
\label{eq:Jc}
\eea
At this point, it is common to further approximate the rather
complicated integration over the angle between $\vec p$ and $\vec k$
\cite{gago2010}.  Here we proceed differently using $z$ as an
integration variable to encode the dependence on this angle. Then the
standard angular and radial integrals are replaced by
\be
\int_{0}^{\frac{\Lambda}{p}}d\beta
\int\limits_{\frac{\beta}{1+\beta}}^{\frac{\beta}{|1-\beta|}} dz=
\int_{0}^{1}dz\int\limits_{\frac{z}{z+1}}^{\frac{z}{1-z}} d\beta+
\int_{1}^{\frac{\Lambda}{p}} dz\int\limits_{\frac{z}{z+1}}^{\frac{z}{z-1}} d\beta,
\ee
where $\ba=k/p$. With this trick, the integral over $z$ in
\eq{eq:gapA2} nicely splits into intervals that correspond to the
definition domains of the function $J$ in \eq{eq:Jc}.  A
straightforward calculation then leads to the following complete
result for the Fermi velocity dressing function in the GW
approximation with RPA polarization effects in the Coulomb
interaction:
\be
A(p)=1+\frac{2}{\pi^2N_f g}\left\{
-\left[\pi-2g+c(g)(g^2-1)\right]\ln\frac{p}{\La} +f(g) +(\pi+4 g)\ln
2-4g \right\}, 
\label{eq:Asolfull}
\ee
where the physical cutoff $\Lambda $ is determined by the size of the
Brillouin zone. The logarithmic term comes from the leading one-loop
correction and was obtained in Ref.~\cite{gonzalez1999} already.  Here
we include all nonsingular terms which were not given therein. They
can be expressed in terms of the function
\bea
f(g)&=&-2(g^2-1)\left\{
g \sum_{n=1}^{\infty}
\frac{(g^2-1)^{n-1}\nu_n-\delta_n}{g^{2n}}
+
\frac{1}{\sqrt{g^2-1}}\left[\mathrm{arccosh}
\,g\ln\left(2\frac{\sqrt{g^2-1}}{g^2}\right)
\right.\right.
\nonumber\\
&&\left.\left.-\pi\arctan\frac{\sqrt{g^2-1}-1}{\sqrt{g^2-1}+1}\right]
\right\} .
\eea
The sum in first term on the right, involving the coefficients
\bea
\de_n=\int_{0}^{1} dz\, (1-z^2)^{n-3/2}\,\arccos z \quad \mbox{and } \;
\nu_n=\int_{0}^{1} dt\, t (1-t^2)^{n-3/2}\, \arccos \frac{1}{t},
\eea
can be evaluated numerically. The full GWA result (\ref{eq:Asolfull})
for the Fermi velocity renormalization function at various values of
the bare effective coupling $\alpha = e^2/\varepsilon v_f $ is shown
in \fig{fig:A-crit-pct}. The logarithmic singularity at $p^2
\rightarrow 0$ is rather weak at weak coupling and the nonsingular
part of the radiative corrections remains small as well. This changes
when $\alpha \approx 2$ is used for suspended graphene. Both
logarithmic and regular contributions are considerably larger and a
correspondingly strong renormalization of the Fermi velocity
results. If we simply use $v_f A(\mu/v_f)$ to estimate the
renormalized Fermi velocity at nonzero chemical potential $\mu$ for a
finite charge-carrier density this result describes the recent
suspended-graphene experiments \cite{elgo2012} so well that a more
quantitative comparison maybe worthwhile \cite{inprep}. The large
effect also shows that a nontrivial dressing function $A(p)$ should be
included in the analysis of the semimetal-insulator transition at
strong coupling, instead of using $A = 1$ as in \cite{gago2010}. We
will turn to this analysis next.

\begin{figure}[t]
%\vspace{0.5cm}
\centering\includegraphics[width=0.6\linewidth]{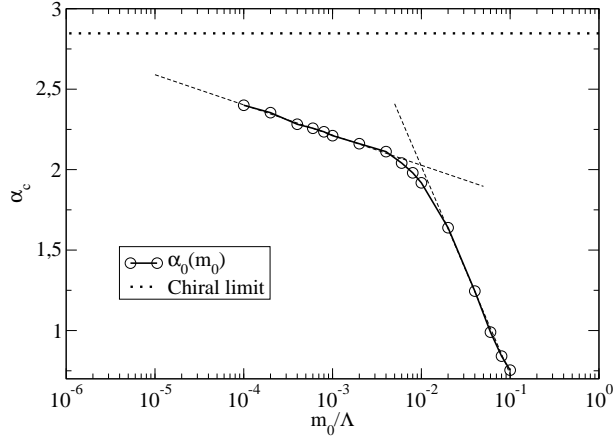}
\caption{\label{fig:mass}
Mass dependence of the critical coupling for $N_f=2$. }
\end{figure}
Having calculated the function $A$ in GWA, we now use this result in
the bifurcation analysis of the gap equation around the critical
point. At leading order in $\Delta $ the gap equation then reduces to:
\be
\Delta(p)=\frac{e^2}{v_F \varepsilon}\int\frac{d^2\vec
  k}{(2\pi)^2}\frac{1}{k\, |\vec p- \vec k|}
\frac{\Delta(k)}{A(k)}
J(z A(k),g). \label{eq:gapD2}
\ee
We proceed by inserting the Fermi velocity dressing function
\eq{eq:Asolfull} into the above equation, and solve the resulting
equation by using standard numerical methods. Note that in this case,
not only the logarithmic but also the regular contributions to $A$
become important at strong coupling.

For the critical coupling we then obtain $\alpha_c=2.85$ which is
larger than the bare coupling $\alpha_0=2.19$ with $\varepsilon =1$
for suspended graphene. This suggests that the Coulomb interactions
might not be strong enough to generate a gapped phase in the
electronic spectrum of suspended graphene. Our result is thus in
contrast to the many previous studies mentioned in the introduction.
The values predicted there all lie below the bare coupling for
suspended graphene. In particular, our results are directly comparable
to the work of Gamayun \emph{et al.}~in Ref.~\cite{gago2010} in which
the same gap equation with the same RPA improved Coulomb interaction
was solved, however, with $A=1$, i.e., with neglecting Fermi velocity
renormalization effects, leading to $\alpha_c=0.92$.

As it turns out, lattice simulations at present also appear to
underestimate the value of the critical coupling
\cite{drula2009,Buividovich:2012uk}. Hence, in order to mimic the
lattice setup, where a non-zero fermion mass term must be used in a
finite volume,\footnote{Finite-volume effects must also be studied
carefully on the lattice. Here QED$_3$ serves as an important example
where it has been shown what a significant impact these can have, see
for example \cite{Goecke:2008zh} and the references therein.}  we have
introduced such an explicitly symmetry breaking term as a seed in the
gap equation (\ref{eq:gapD2}) here as well, and calculated the
corresponding pseudocritical couplings.  Analyzing the results
displayed in \fig{fig:mass}, we notice that the diagram contains two
distinct regions, each of them displaying a logarithmic dependence of
the critical coupling on the bare mass (dashed lines in the
figure). The scale separating the two regions, where the dynamically
generated mass is bigger or smaller than the bare fermion mass, is of
the order of magnitude $m_0/\Lambda=10^{-2}$. Note however that a
proper comparison with the lattice requires taking into account the
different UV scales, as graphene lattice simulations have been
performed on both hexagonal \cite{Buividovich:2012uk} and square
\cite{drula2009} lattices.  Further studies will be necessary to
clarify the issue.

\section{Summary and outlook}\label{sec5}

In this work, we have provided an estimate of the critical coupling
for the semimetal-insulator phase transition in graphene. We have
employed the Dyson-Schwin\-ger equation for the fermion propagator in
the Dirac cone approximation with a bare vertex and Coulomb
interactions including frequency dependent polarization effects at the
RPA level. Importantly, however, we have explicitly taken into account
the GWA running of the Fermi velocity. In this setting, we have
obtained a critical coupling of $\al_c = 2.85$, which is larger than
the bare coupling $\al_0=2.19$ for suspended graphene. This allows us
to conclude that the logarithmic growth of the Fermi velocity near
charge neutrality weakens the Coulomb interaction preventing the
emergence of a gapped phase, in agreement with the experimental
observations \cite{elgo2012}.

Despite the qualitative agreement with experimental observations, a
complete understanding of the physical picture calls for a full
nonperturbative investigation, i.e., to solve the coupled system of
integral equations for the photon and fermion propagators
simultaneously. It is likely that including the nonperturbative gauge
propagator and improving the vertex functions may lead to further
corrections of the critical coupling. We recall that nonperturbative
contributions to both the photon propagator and vertex functions lead
to significant corrections in ordinary QED$_3$, of the order of twenty
percent for the photon propagator and ten percent for the vertex
\cite{fial2004,Braun:2010qs}. These effects, along with finite volume
and mass corrections, might in the end together explain the
discrepancy between experiment and lattice simulations
\cite{drula2009,Buividovich:2012uk}. 

Moreover, induced four-fermion contact interactions might become
important at strong coupling as well. Such local four-fermion
interaction terms as allowed by the symmetries should be considered,
e.g., to model electron-phonon interactions, in addition to the
long-range Coulomb fields in the effective low-energy continuum theory
for the electronic excitations of graphene in the future.\footnote{For
first studies in this direction we refer the reader to
Refs.~\cite{gago2010,Mesterhazy:2012ei}.} Finally, another interesting
problem is the inclusion a chemical potential for a finite density of
charge carriers as induced experimentally, e.g., by chemical
doping. While the fermion-sign problem makes this problem hard to
address with lattice simulations, the modifications to the
Dyson-Schwinger equations are quite straightforward. The resulting
effects of a finite charge-carrier density on the running Fermi
velocity are the subject of an upcoming publication.

\vspace{0.5cm}

\leftline{\bf Acknowledgments}

We would like to thank P.~Buividovich and M.~Polikarpov for valuable
discussions. This work was supported by the Deutsche
Forschungsgemeinschaft within the SFB 634, the Helmholtz International
Center for FAIR within the LOEWE program of the State of Hesse, and
the Helmholtz Young Investigator Group No.~VH-NG-332.

%\bibliography{$HOME/bibliography/biblio}
%\ifgeneratebbl
%\bibliographystyle{utphys}
%\bibliography{$HOME/bibliography/biblio}
%\else
%\input{confX-carina.bbl}
%\fi

\providecommand{\href}[2]{#2}\begingroup\raggedright\endgroup

\end{document}